\documentclass[prx,aps,amsfonts,twocolumn, groupedaddresses,nofootinbib,superscriptaddress]{revtex4-2}

\usepackage{graphicx} % Required for inserting images
\usepackage{amsmath, amssymb, amsfonts, mathtools}
\usepackage[colorlinks=true, urlcolor=violet, linkcolor=blue, citecolor=red, hyperindex=true, linktocpage=true]{hyperref}
\usepackage{xcolor, color, physics,overpic}
\usepackage{comment}

\def \beq { \begin{equation}}
\def \eeq {\end{equation}}

\begin{document}

\title{
Planckian bound on the local equilibration time
}
\author{Marvin Qi}
%\email{marvinqi@uchicago.edu}
\affiliation{Leinweber Institute for Theoretical Physics \& James Franck Institute, University of Chicago}%, Chicago, IL 60637, USA}

\author{Alexey Milekhin}
%\email{milekhin@uky.edu}
\affiliation{Department of Physics and Astronomy, University of Kentucky}

\author{Luca Delacr\'etaz}
%\email{lvd@uchicago.edu}
\affiliation{Leinweber Institute for Theoretical Physics \& James Franck Institute, University of Chicago}

\begin{abstract}
The local equilibration time $\tau_{\rm eq}$ of quantum many-body systems is conjectured to be bounded below by the Planckian time $\hbar /T$. We formalize this conjecture by defining $\tau_{\rm eq}$ as the time scale at which a hydrodynamic description emerges for conserved densities. Drawing on analytic properties of real time thermal correlators, we establish a rigorous lower bound $\tau_{\rm eq} \geq \alpha  \hbar /T$ on the onset of hydrodynamic behavior in a `regulated' thermal two-point function.  The dimensionless coefficient $\alpha $ depends only on dimensionality and the type of hydrodynamic or diffusive behavior that emerges, and is independent of the thermalization mechanism or other microscopic details. This bound applies universally to local quantum many-body systems, with or without a quasiparticle description, including in the presence of inelastic scattering. 

\end{abstract}

\date{\today}

\maketitle

%######################################################################%
%======================================================================%
%======================================================================%
%======================================================================%
%######################################################################%
\section{Introduction}

The relevance of the Planckian timescale $ {\hbar} / {T}$ in quantum statistical physics has long been appreciated \cite{Planck1901,peierls1934remarks,Matsubara1955}. It has gathered attention recently due to the possibility that it universally bounds the local equilibration time of quantum many-body systems \cite{qptbook,Zaanen2004,Hartnoll:2021ydi}
\begin{equation}\label{eq_Planckian_conjecture}
\tau_{\rm eq} \gtrsim \tau_{\text{Pl}} \equiv \frac{\hbar}{T} \, .
\end{equation}
Theoretically, this bound is motivated by the observation that while thermalization is parametrically slow $\tau_{\rm eq}\sim 1/\lambda^2$ at weak coupling $\lambda\ll 1$, many tractable strongly coupled systems thermalize at the Planckian rate $\tau_{\rm eq}\sim \tau_{\rm Pl}$. This conjecture was also originally motivated by the linear-in-temperature resistivity seen across materials in the normal state of high-temperature superconductors, see \cite{Hartnoll:2021ydi} for a review.

A bound of the form \eqref{eq_Planckian_conjecture} may seem natural from the perspective of the Heisenberg energy-time uncertainty relation $\Delta E \Delta t \gtrsim \hbar$, postulating $\Delta E = T$ due to thermal fluctuations. However, making this intuition precise and establishing \eqref{eq_Planckian_conjecture} for a sharply defined $\tau_{\rm eq}$ has proven challenging. In this work, we will adopt the proposed universal definition of $\tau_{\rm eq}$ as the onset time of hydrodynamic behavior \cite{Delacretaz:2023pxm}. Hydrodynamics, broadly viewed as the slow dynamics of conserved densities, emerges in essentially any local many-body system, from spin chains, to metals or insulators, and quantum field theories. This definition is thus broadly applicable, and is insensitive to the mechanism of how thermalization occurs in the system. 

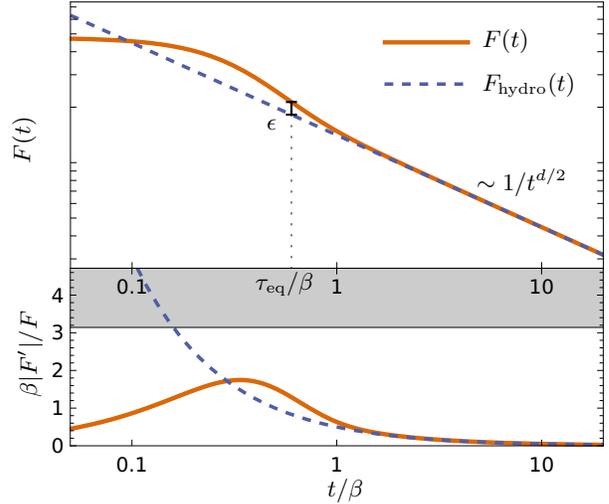
\begin{figure}[h]
\centerline{
\begin{overpic}[width=0.85\linewidth,tics=10,trim={-0.15cm 0cm 0.07cm 0cm},clip]{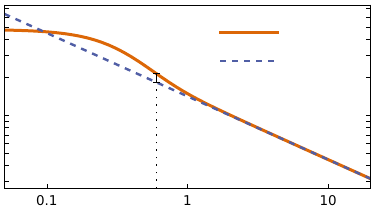}
	 \put (-7,25) {\rotatebox{90}{$F(t)$}} 
	 \put (78,46) {$F(t)$} 
	 \put (78,38) {$F_{\rm hydro}(t)$} 
	 \put (39,31) {$\epsilon$} 
	 \put (37,2) {$\tau_{\rm eq}/\beta$} 
	 \put (77,20) {$\sim 1/t^{d/2}$} 
\end{overpic}
}
\vspace{-15pt}
\centerline{
\begin{overpic}[width=0.85\linewidth,tics=10,trim={0.0cm 0cm 0cm 0cm},clip]{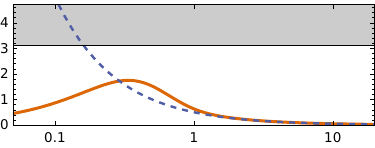}
	 \put (-6,15) {\rotatebox{90}{$\beta |F'|/F$}} 
	 \put (50,-3) {$t/\beta$} 
\end{overpic}
}
\caption{\label{fig_Fig1}Top: the local equilibration time $\tau_{\rm eq}$ can be defined as the time beyond which correlation functions (orange) agree with hydrodynamic predictions (blue dashed). Bottom: analyticity of the two-sided correlator $F(t) = \Tr [\sqrt{\rho}n(t)\sqrt{\rho}n]$ implies that its derivative is bounded $\beta |F'|/F \leq \pi$ (gray forbidden region). Hydrodynamic behavior cannot onset too early or it would violate that bound.}
\end{figure} 

The expectation, which is bourne out at weak coupling, is that a hydrodynamic description emerges more quickly as interactions become stronger. This leads to the question: by tuning interactions to be arbitrarily large, can hydrodynamics emerge arbitrarily quickly? It was shown in \cite{Delacretaz:2023pxm} that this is not the case. Hydrodynamic fluctuations produce power-law corrections to the asymptotic behavior of observables; these become large at earlier times, at which point fluctuating hydrodynamics becomes uncontrolled. $\tau_{\rm eq}$ is then bounded below by this strong coupling scale of hydrodynamics. This was sufficient to establish \eqref{eq_Planckian_conjecture} for relativistic quantum field theories.

The goal of this paper is to prove a Planckian bound \eqref{eq_Planckian_conjecture} for general quantum many-body systems. The Planckian time of course  makes a conspicuous appearance in imaginary time observables. By analytic continuation, it is therefore natural to expect it to matter in real time data as well. Analytic properties of thermal correlators were used in Ref.~\cite{Maldacena:2015waa} to establish a bound on the Lyapunov exponent $1/\lambda_L \geq \hbar /(2\pi T)$ that appears in the out-of-time-ordered correlators of certain large-$N$ or semiclassical systems. As we will show, similar techniques can be used to prove bound the rate of change of a specific `regularized' two-point correlation function. This bound will then constrain the onset time of hydrodynamic behavior in this correlator.

Planckian bounds have been proposed from several different perspectives in the past. Our approach bounds the emergence of hydrodynamics, so that locality of the quantum many-body system plays a central role. Other notions of thermalization allow for a formulation of the Planckian conjecture where locality is less central \cite{Lucas:2018wsc,Abiuso:2025baf}. The inequality \eqref{eq_Planckian_conjecture} can also be interpreted as a bound on how strongly interacting a quantum many-body system can be \cite{doi:10.1073/pnas.2216241120}. See \cite{Reimann_2016,Nussinov:2021fgc,Pappalardi:2021ahe,Chowdhury:2026rxn} for further discussions on Planckian-like bounds. Our work also contributes to hopes of constraining transport and thermalization away from weak coupling through universal bounds \cite{Kovtun:2004de,Buchel:2007mf,Hartnoll:2014lpa}. Progress has been made using causality constraints \cite{Baier:2007ix,Hartman:2017hhp,Delacretaz:2021ufg,Heller:2022ejw, Chowdhury:2025qyc}; sum-rules may also produce bounds, although these typically constrain equilibrium observables \cite{Romatschke:2009ng,Allameh:2024qqp,Chowdhury:2025dlx,Chowdhury:2025tzq}.

%######################################################################%
%======================================================================%
%======================================================================%
%======================================================================%
%######################################################################%
\section{Local equilibration time and hydrodynamics}

The onset of hydrodynamic behavior provides a universal framework to define the local equilibration time $\tau_{\rm eq}$ \cite{Delacretaz:2023pxm}. There are several ways to make this definition sharp: a fairly intuitive one is to let $\tau_{\rm eq}$ be the time scale beyond which a simple correlation function agrees, up to a small error, with the hydrodynamic prediction. In this paper we will consider the following correlator of densities:
\begin{equation}\label{eq:Fnn_def}  
F_{nn}(t,x) \equiv \Tr \left( \sqrt{\rho} n(t,x) \sqrt{\rho} n \right)\, , 
\end{equation}
where $\rho = e^{-\beta H}/\Tr(e^{-\beta H})$ is the thermal density matrix, and $n$ is a conserved density satisfying a continuity equation $\dot n + \nabla\cdot j = 0$. We consider this somewhat unusual object, sometimes referred to as the `two-sided correlator', because we will see it lends itself well to establishing a Planckian bound on $\tau_{\rm eq}$. However, it is closely related to more familiar (and measurable) observables such as the response functions or dynamic structure factor, see Eq.~\eqref{eq_GS_GA_1} below.

At late times, hydrodynamics predicts that the correlation function \eqref{eq:Fnn_def} behaves as
\begin{equation} \label{eq:Fcl}
    F_{nn}(t,x) \xrightarrow{t \gg \tau_{\rm eq}} F_{\rm hydro} (t,x) = \frac{\chi T}{(4\pi D t)^{d/2}} e^{ - \frac{x^2}{4Dt} }\, ,
\end{equation}
where $D$ is the diffusivity, $\chi = dn/d\mu$ the charge susceptibility, and $d$ the spatial dimension.
Let us now set $x=0$ and consider $F_{nn}(t) \equiv F_{nn}(t,0)$.
We define $\tau_{\rm eq}(\epsilon)$ to be the smallest time $\tau$ such that, for any $t > \tau$, the relative error between $F_{nn}(t)$ and $F_{\rm hydro}(t)$ (and their derivative) is less than $\epsilon$. Formally, $\tau_{\rm eq}(\epsilon)$ is defined as 
\begin{widetext}
\begin{equation}\label{eq_taueq_def}
    \tau_{\rm eq}(\epsilon) \equiv \inf \Bigg\{\tau : \sup_{t>\tau}  \left\{ \left| \frac{F_{nn}(t) - F_{\rm hydro}(t)}{F_{\rm hydro}(t)} \right|,  \; 
    \left| \frac{F'_{nn}(t) - F'_{\rm hydro}(t)}{F'_{\rm hydro}(t)} \right|  \right\} < \epsilon\Bigg\}.
\end{equation}
\end{widetext}
When it will not cause confusion, we will omit the $\epsilon$ dependence of $\tau_{\rm eq}$.

We will see that this time scale has many of the properties that could be desired from a Planckian bound. 
\begin{enumerate}
\item[(a)]
	It can be defined away from weak coupling.
\item[(b)]
	For weakly coupled quasiparticle systems, it agrees parametrically with the large-angle scattering time.
\item[(c)]
	It satisfies a Planckian bound $\tau_{\rm eq} \geq \alpha\frac{\hbar}{T}$, with coefficient $\alpha$ determined below.
\item[(d)]
	The Planckian bound is only saturated for strongly correlated systems.
\end{enumerate}
Property (a) is clear from the definition, and (b) follows from the fact that the (large-angle) scattering time sets the  scale of onset of hydrodynamic behavior in weakly interacting quantum many-body systems. Deriving (c) is the subject of this paper. The final property (d) is more subtle: we expect it to be true (but cannot prove it) for clean quantum many-body systems. However, it is not true in general with this definition of $\tau_{\rm eq}$: we will see that a free particle in a disordered landscape exhibiting Brownian motion saturates our bound. This is an aspect of our results which could be improved upon, and will be further discussed in the conclusions.

The Planckian bound will be proven in the next section, and follows from analyticity properties of $F_{nn}(t)$ in the complex plane. Before providing the rigorous proof, we spell out the result here. The bound is simplest in situations where the symmetric charge Green's function is positive for all times
\begin{equation}\label{eq_GS_positive}
G^S_{nn}(t) \equiv \frac{1}{2} \Tr \left(\rho \{n(t),n(0)\}\right) > 0 \, , \qquad \forall t\, .
\end{equation}
This holds in a number of examples 
(and is often observed numerically), but is not true in general: we make this additional assumption for pedagogy and will lift it later. When \eqref{eq_GS_positive} holds, we show in the next section that the rate of change of $F_{nn}(t)$ is bounded:
\begin{equation}\label{eq_FprimeOverF}
\left| \frac{F_{nn}'(t)}{F_{nn}(t)} \right| \leq \pi\frac{T}{\hbar}\, .
\end{equation}
Clearly, this inequality is not satisfied by the hydrodynamic form of the correlator \eqref{eq:Fcl} at early times. This is not a contradiction, since hydrodynamics does not emerge at arbitrarily early times, but will set a bound on the onset of hydrodynamic behavior. Indeed, Eq.~\eqref{eq_taueq_def} then implies that for any time $t>\tau_{\rm eq}(\epsilon)$, we have
\begin{equation}
     \frac{d}{2t}  \leq \pi\frac{T}{\hbar} + \mathcal{O}(\epsilon),
\end{equation}
leading to the bound
\begin{equation}\label{eq_bound_simple}
\tau_{\rm eq} \geq \frac{d}{2\pi } \frac{\hbar}{T}\, .
\end{equation}
This is a Planckian bound on the emergence of diffusion in $d$ spatial dimensions. Other hydrodynamic behavior (sound modes, subdiffusion or superdiffusion) can be similarly bounded, as will be discussed in Sec.~\ref{sec_other}. It is interesting that the dimension $d$ enters in the bound -- we will further discuss this in the conclusion.

In the next section, we show that \eqref{eq_FprimeOverF} follows from analyticity when \eqref{eq_GS_positive} holds. We then lift the assumption \eqref{eq_GS_positive} and derive a weaker but more general Planckian bound.

%######################################################################%
%======================================================================%
%======================================================================%
%======================================================================%
%######################################################################%
\section{Analytic structure of correlators} \label{sec:analyticity}

To derive a general bound on two-point thermal correlation functions, we follow the approach introduced by Maldacena, Shenker, and Stanford (MSS) in \cite{Maldacena:2015waa}. Consider the following correlator of two Hermitian operators $A$ and $B$:
\begin{align}
F_{AB}(t+i\tau) 
	&= \Tr \left(\sqrt\rho A(t+i\tau) \sqrt\rho B \right)\, , 
\end{align}
with $A(z) \equiv e^{iHz} A e^{-iHz}$ (we set $\hbar=1$ in what follows). The spectral representation of $F_{AB}$ shows that it is an analytic function of $z=t+i\tau$ in the strip $|\Im(z)| = |\tau|  \leq \beta/2$. When $z=t$ is real, $F_{AB}(t)$ is real, due to Hermiticity of $A$ and $B$, and corresponds to the `two-sided correlator' introduced in \eqref{eq:Fnn_def}. In what follows, we will derive a bound on the rate of change of $F_{AB}(t)$. Before doing so, however, we comment below on the interpretation of the correlation function $F_{AB}(t)$ and its relation to more conventional physical observables.

Analyicity of $F_{AB}(z)$ encodes the time-domain fluctuation-dissipation relation, as observed in \cite{Pappalardi:2021ahe}. To see this, note that the real and imaginary parts of $F_{AB}(z)$ at $\Im(z) = \beta/2$ correspond to the symmetric and antisymmectric Green's functions
\begin{align}\label{eq_GS_GA_1}
    G^S_{AB}(t) &= \frac{1}{2} \Tr \left( \rho \{ A(t), B \} \right) = \Re(F_{AB}(t + i \beta/2)) \\
    G^A_{AB}(t) &= \frac{1}{2i} \Tr \left( \rho [A(t), B] \right) = \Im(F_{AB}(t + i \beta/2))
\end{align}
respectively. Analyticity implies that $F_{AB}(t + i \beta/2) = {\exp} (i \frac{\beta}{2} \partial_t{)} F_{AB}(t)$, and thus $G^S_{AB}(t) = {\cos} ( \frac{\beta }{2} \partial_t ) F(t)$ and $ G^A_{AB}(t) = {\sin} (  \frac{\beta }{2} \partial_t ) F(t)$,
which gives the time-domain fluctuation-dissipation relation between $G^S_{AB}$ and $G^A_{AB}$.

We now proceed to derive a bound on the growth of $F_{AB}(t)$. We recall the key fact from complex analysis that was used by MSS to derive a bound on the quantum Lyapunov exponent. Let $f(z)$ be a function with the following properties: 
\begin{enumerate}
    \item $f(t + i \tau)$ is analytic in the half strip $t > t_0$ and $|\tau| \leq \beta/2$.
    \item $f(t)$ is real.
    \item $|f(t+i \tau)| \leq 1$ in the half-strip $t > t_0$ and $|\tau| \leq \beta/2$.
\end{enumerate}
Then $f(t)$ obeys the inequality 
\begin{equation} \label{eq:fbound}
    \frac{1}{1-f} \left| \frac{df}{dt} \right| \leq \frac{\pi}{\beta} \coth \left( \frac{\pi}{\beta} (t-t_0) \right).
\end{equation}
This follows immediately from the Schwarz-Pick theorem after mapping the half strip to the disk, see Section 4.1 of Ref.~\cite{Maldacena:2015waa} for details. Note that we do not make use of the KMS condition $f(t) = f(-t)$.

The strategy will be to apply the above inequality to the function 
\begin{equation} \label{eq:fdef}
    f(z) = 1 - \frac{F_{AB}(z)}{C}
\end{equation}
for a (yet to be specified) constant $C>0$. This leads to the bound
\begin{equation} \label{eq:bound1}
    \frac{1}{F_{AB}} \left| \frac{d F_{AB}}{dt} \right| \leq \frac{\pi}{\beta} \coth \left( \frac{\pi}{\beta} (t - t_0) \right). 
\end{equation}
Properties $1$ and $2$ follow immediately from analyticity of $F_{AB}(z)$ and reality of $F_{AB}(t)$. Property 3 will require $F_{AB}(z)$ to satisfy additional conditions which follow from hydrodynamics, as we now explain.

Hydrodynamics controls the late time behavior of correlation functions. Two aspects of hydrodynamic predictions will be useful in establishing property 3 above. First, $G^S(t)\gg G^A(t)$ at late times, or more precisely
\begin{equation} \label{eq:classical}
    \frac{G^A_{AB}(t)}{G^S_{AB}(t)} = O \left( \frac{\beta \hbar}{t} \right) \quad \text{ as } t \to \infty\, .
\end{equation}
This applies to any operators $A$, $B$ overlapping with the conserved density. Loosely speaking, effectively such operators approximately commute at late times. The second useful fact is that $\Re F_{AB}(z)$ is positive in the hydrodynamic regime $\Re z \to \infty$ (it is given by \eqref{eq:Fcl}), and therefore there exists $t_0 < \infty$ such that
\begin{equation}\label{eq_t0}
\Re F_{AB}(z) > 0 \, , \quad \forall \Re z\geq t_0\, .
\end{equation}
Eqs.~\eqref{eq:classical} and \eqref{eq_t0} are not generally true: they are predictions of hydrodynamics. If either were violated, hydrodynamics would never emerge and the Planckian bound would trivially hold, with $\tau_{\rm eq} = \infty$. We therefore assume Eqs.~\eqref{eq:classical} and \eqref{eq_t0} in what follows.

With these assumptions on $F_{AB}(z)$ at hand, we now prove that $F_{AB}(t)$ satisfies property 3. We follow the procedure for bounding $|f(z)|$ outlined in \cite{Maldacena:2015waa}: first, we show that $|f(z)| \leq 1$ at the boundaries of the half-strip, \emph{i.e.} at $\Re(z) = t_0$ (the left boundary) and at the top and bottom boundaries $\Im(z) = \pm \beta/2$; next, we show that $|f(z)|$ is bounded by a constant everywhere in the interior. It then follows, via the Phragm\'en-Lindel\"of principle, that $|f(z)| \leq 1$ everywhere in the half-strip, proving property 3. 

Below, we will make frequent use of the fact that $|F_{AB}(z)|$ is bounded on the strip $\Re z \geq t_0$. This holds both in quantum field theory, and for local lattice Hamiltonians. For example, for any lattice model with finite dimensional Hilbert spaces at each site, one can show (see Appendix \ref{app:lattice_bound})
\begin{equation}
\label{eq:lattice_bound}
    |F_{AB}(z)| \leq  \Vert A \Vert \, \Vert B \Vert,
\end{equation}
where $||\cdot||$ is the operator infinity-norm, i.e. the maximal magnitude of an eigenvalue.

Let us first consider the behavior of $F_{AB}(z)$ at the top and bottom boundaries of the half-strip, at $|\Im(z)| = \beta/2$. The condition $|f(z)| \leq 1$ is equivalent to 
\begin{equation} \label{eq:Fcondition}
    \left| \frac{F_{AB}(z)}{C} \right|^2 \leq 2 \Re \left( \frac{F_{AB}(z)}{C} \right).
\end{equation}
At the boundary of the strip, $F_{AB}(t + i \frac{\beta}{2}) = G^S_{AB}(t) + i G^A_{AB}(t)$. Since $C$ is real and positive, the condition becomes 
\begin{equation} \label{eq:Gcondition}
    \left|G^S_{AB}(t)\right|^2 + \left|G^A_{AB}(t)\right|^2 \leq 2 C G^S_{AB}(t).
\end{equation} 
We see that the inequality is satisfied if $C$ is chosen as
\begin{equation} \label{eq:C1}
    C \geq \max_{t\geq t_0}  \left( G^S_{AB}(t), \frac{ \left|G^A_{AB}(t)\right|^2 }{G^S_{AB}(t)} \right). 
\end{equation} 
$G^S_{AB}(t)$ is bounded above by $|F_{AB}(t + i \frac{\beta}{2})|$, which is bounded by \eqref{eq:lattice_bound}. On the other hand, $\frac{|G^A_{AB}(t)|^2}{G^S_{AB}(t)}$ approaches zero as $t \to \infty$ by \eqref{eq:classical}, and doesn't diverge anywhere since $G^S_{AB}(t) > 0$ for all $t \geq t_0$ by \eqref{eq_t0}. Therefore it reaches a maximum at some value of $t$. Choosing a $C$ such that \eqref{eq:C1} is satisfied is therefore always possible. 

Next, we show that $|f(z)| \leq 1$ on the left boundary, at $\Re(z) = t_0$. 
From \eqref{eq:Fcondition}, we see that $|f(z)| \leq 1$ is satisfied as long as 
\begin{equation} \label{eq:C2}
    C \geq \sup_\tau \frac{|F_{AB}(t_0 + i \tau)|^2}{2 \Re (F_{AB}(t_0+i \tau))}.
\end{equation}
Because the left boundary $\Re(z) = t_0$ is a compact interval, the right-hand side is finite as long as the denominator is nonzero everywhere on the interval. Since this was assumed in \eqref{eq_t0}, finiteness immediately follows. We choose $C$ to be large enough such that \eqref{eq:C1} and \eqref{eq:C2} are satisfied.

To finish the proof of property 3, we now only need to bound $|f(z)|$ by a constant everywhere in the interior of the half-strip. This follows immediately from boundedness of $F_{AB}(z)$ in \eqref{eq:lattice_bound}. Since we have already bounded $|f(z)| \leq 1$ on the boundary of the half-strip, the Phragm\'en-Lindel\"of principle then implies that $|f(z)| \leq 1$ everywhere in the half-strip, establishing property 3. This concludes the proof of the bound \eqref{eq:bound1} on $F_{AB}(t)$. 

This bound can be improved if $G^S_{AB}(t) > 0$ for all $t$ and $A$ and $B$ are bounded operators. In this case, since $\Re(F_{AB}(z))$ is a bounded harmonic function and is positive on the boundary of the strip $|\Im(z)| \leq \frac{\beta}{2}$, it follows that $\Re(F_{AB}(z))$ is positive everywhere in the interior. In other words, $t_0$ in \eqref{eq_t0} can be taken to $t_0 \to -\infty$. When $G^S_{AB}(t) > 0$, we therefore obtain the stronger inequality 
\begin{equation}
    \frac{1}{F_{AB}} \left| \frac{d F_{AB}}{dt} \right| \leq \frac{\pi}{\beta} 
\end{equation}
used to derive the Planckian bound \eqref{eq_bound_simple} in the previous section. In Appendix \ref{app:weakening_assumptions}, we relax the condition $G^S_{AB}(t) > 0$ and obtain a looser bound on $\tau_{\rm eq}$.

%######################################################################%
%======================================================================%
%======================================================================%
%======================================================================%
%######################################################################%
\section{Effective field theory interpretation}

Fluctuating hydrodynamics is quantitatively described by an Effective Field Theory (EFT) that systematically captures irrelevant corrections to a given hydrodynamic behavior, say diffusion \cite{Martin:1973zz,dominicis1976techniques,Janssen:1976qag,Crossley:2015evo,Haehl:2015foa,Jensen:2018hse,Chen-Lin:2018kfl,Michailidis:2023mkd,Delacretaz:2023pxm}. For example, the density correlator takes the form
\begin{equation}\label{eq_Fnn_hydro}
F_{nn}(t) = 
    \frac{\chi T}{(4\pi D t)^{d/2}} \left[1 + \left(\frac{\tau_{\rm loop}}{t}\right)^{d/2} + \frac{\tau_{\rm deriv}}{t} + \cdots\right]\, , 
\end{equation}
where $\tau_{\rm loop}$ arises from the leading (1-loop) fluctuation corrections, and $\tau_{\rm deriv}$ from the leading higher-derivative corrections. These intermediate time corrections give microscopics-independent insight into how quickly hydrodynamic behavior can emerge. The loop (or strong-coupling) scale of diffusive EFTs can be obtained from the EFT itself and was argued to bound equilibration in Ref.~\cite{Delacretaz:2023pxm} $\tau_{\rm eq} \geq \tau_{\rm loop} \propto \frac{(\chi T)^{2/d}}{D} \left(\frac{dD/dn}{D}\right)^{4/d}$. What is the scale of higher-derivative corrections, $\tau_{\rm deriv}$? While these are difficult to constrain in general from the EFT alone, a subset of higher derivative corrections are enforced by KMS symmetry \cite{Sieberer:2015hba,Crossley:2015evo,Haehl:2015foa,Jensen:2018hse}. Consider for example the following terms in the diffusion EFT:
\begin{align}\label{eq_SEFT_ex}
&S[\mu,\phi_a]=\\ \notag
&\chi \int \mu (\partial_t + D \nabla^2)\phi_a 
    + iTD \nabla\phi_a \frac{\tfrac12\beta\partial_t}{\tan \tfrac12\beta\partial_t} \nabla\phi_a + \cdots\, .
\end{align}
KMS symmetry requires the same coefficient $D$ to appear in front of two different terms, with a specific derivative structure. 
This action reduces to the classical MSR description \cite{Martin:1973zz,dominicis1976techniques,Janssen:1976qag} to leading order in derivatives $\frac{\frac12\beta\partial_t}{\tan \frac12\beta\partial_t} \to 1$. This example suggests that the scale for higher-derivative terms is at least Planckian, $\tau_{\rm deriv}\geq \beta$, thus making our Planckian bound \eqref{eq_bound_simple} appear to be natural from the EFT perspective. 

However, this observation is too naive: the EFT in general contains many other higher derivative terms (the $\cdots$ in \eqref{eq_SEFT_ex}), which can cancel the contributions from \eqref{eq_SEFT_ex}. Therefore, the EFT predictions do not automatically satisfy the constraints from the previous section. This is because imposing KMS does not guarantee analyticity on the strip: analyticity leads to an additional UV/IR constraint on the Wilsonian coefficients of the EFT. This is further discussed in App.~\ref{app_superplanckianhydro}.

%######################################################################%
%======================================================================%
%======================================================================%
%======================================================================%
%######################################################################%
\section{Bounding other late time behavior}\label{sec_other}

While we focused on diffusion for simplicity, our approach similarly bounds the emergence of any hydrodynamic behavior. For example, the density correlator can be generalized to 
\begin{equation}
F(t) \sim \frac{1}{t^{d/z}}
\end{equation}
to describe superdiffusion ($z<2$) or subdiffusion ($z>2$) at late times. The Planckian bound Eq.~\eqref{eq_bound_simple} then becomes
\begin{equation}\label{eq_bound_subdiffusion}
\tau_{\rm eq} \geq \frac{d}{z\pi } \frac{\hbar}{T}\, ,
\end{equation}
if Eq.~\eqref{eq_GS_positive} is assumed (this assumption can again be lifted by slightly worsening the bound using the method in App.~\ref{app:weakening_assumptions}). Our approach also applies to sound modes, with minor modifications to follow the sound front which is no longer at $x=0$: one considers the spatially resolved correlator $F(t,x) = \Tr (\rho^{1/2} n(t,x) \rho^{1/2} n)$, and then sets $x=v_{\rm sound} t$ at the end of the calculation.

Beyond hydrodynamics, another behavior that is often observed in the late time limit of thermal correlators is exponential decay
\begin{equation}\label{eq_F_exp}
F(t) \simeq Ce^{-\Gamma t} + \cdots \, , \qquad \hbox{as } t\to \infty\, .
\end{equation}
While exponential decay is not generic---there are many models for which no local operator decays exponentially, due to hydrodynamic tails \cite{Delacretaz:2023pxm}---in certain situations a symmetry can forbid overlap with hydrodynamic fields, making exponential decay at least possible. Examples include the $\mathbb Z_2$-odd spin operator in the 2+1d Ising CFT or lattice model. If one assumes exponential decay of such operators, can one bound the exponent $\Gamma$? This is not the case: a counterexample is given by the two-point function of a scalar operator in a 1+1d CFT 
\begin{equation}\label{eq_2dCFT_example}
F(t) = \left(\frac{\pi/\beta}{\cosh \frac{\pi t}{\beta}}\right)^{2\Delta}\, , 
\end{equation}
which at late times decays as \eqref{eq_F_exp} at late times, with $\Gamma = \frac{2\pi}{\beta} \Delta$. Clearly, there can be no upper bound on $\Gamma$. While \eqref{eq_2dCFT_example} is unbounded on the strip, one can alternatively consider the example $F(z) = e^{-\Gamma \sqrt{z^2 + (\beta/2)^2}}$ which is both analytic and bounded on the strip $|\Im z|  \leq \beta/2$.

These examples show that the exponential decay rate of correlators is not expected to be bounded on general grounds without additional conditions. Nevertheless, one could formulate a Planckian conjecture on exponential decay as follows: that the slowest (non-hydrodynamic) decay rate $\Gamma_{\rm min}$---sometimes referred to as `Liouvillian gap'---is bounded by $T/\hbar$. In holographic models, this is a conjecture about the slowest non-hydrodynamic quasi-normal mode. It would interesting to prove this alternative Planckian conjecture, which would require tools beyond those used in this paper. 
We emphasize though that the Liouvillian gap does not generically exist, so that this version of the Planckian conjecture cannot apply to general local many-body systems: it requires the existence of symmetry sectors that are untouched by hydrodynamic tails, and even under these conditions exponential decay is not guaranteed \cite{Dodelson:2024atp,McCulloch:2025fzk}.

%######################################################################%
%======================================================================%
%======================================================================%
%======================================================================%
%######################################################################%
\section{Discussion}

We have shown that the onset of hydrodynamic behavior in the two-sided density correlator $F(t) = \Tr \left( \sqrt{\rho} n(t)\sqrt{\rho} n\right)$ is bounded by the Planckian time 
\begin{equation}\label{eq_bound_conclusion}
\tau_{\rm eq} \geq \alpha \frac{\hbar}{T}\, , 
\end{equation}
with coefficient $\alpha$ depending on the type of hydrodynamic behavior that emerges (see Eqs.~\eqref{eq_bound_simple} and \eqref{eq_bound_subdiffusion}). This bound is parametrically loose at weak coupling---quasiparticles with a small cross section have a large $\tau_{\rm eq} \simeq \frac{1}{n \langle \sigma\rangle v}$---but imposes a tight constraint on the dynamics of strongly coupled systems, where analytic tools are typically less available.

Our result applies broadly to quantum many-body systems; let us comment on its implications for strange metals, which originally motivated the Planckian conjecture \cite{Bruin_Planckian,Zaanen2004}. While we considered a somewhat peculiar `two-sided' correlator to establish a rigorous bound, we expect the emergence of hydrodynamics in conventional real time observables to typically have similar properties. Time-resolved near-field spectroscopy should precisely be able to probe such local observables \cite{Ni2018}, although the Coulomb interaction should be screened for the plasmon dynamics to be diffusive and track the emergence of hydrodynamics \cite{PhysRevLett.117.076805,doi:10.1126/science.aan2735}. In the meantime, frequency resolved observables can offer a proxy for our criterion: the leading low frequency corrections to $\sigma_{\rm dc}$ identify the time scale $\tau_{\rm eq}$. Thus, as emphasized in \cite{Delacretaz:2023pxm}, the observation of $\omega/T$ scaling in the optical conductivity of the cuprates and other strange metals \cite{Sachdev:2011cs,Delacretaz:2016ivq,Michon2023} is possibly the most convincing evidence of Planckian thermalization, namely that these materials are close to saturating our bound Eq.~\eqref{eq_bound_conclusion}. Interestingly, systems that saturate the Planckian bound have strong bounds on their diffusivities due to causality---assuming that correlations cannot spread faster than a velocity $v$---which leads to lower bounds on resistivity \cite{Hartman:2017hhp}
\begin{equation}\label{eq_causality}
D \lesssim v^2 \frac{\hbar}{T} \quad \Rightarrow \quad
\rho_{\rm dc}
	\gtrsim \frac{m_*}{n} \frac{T}{\hbar}\, .
\end{equation}
For causality bound to apply rigorously, $v$ should correspond to an appropriate Lieb-Robinson velocity; we have set it to $v_F$ to obtain a more favorable suggestive bound on the dc resistivity. We see that from this perspective, metals saturating the Planckian bound \eqref{eq_bound_conclusion} are forced to be poor metals with resistivity no {\em smaller} than linear in $T$! For example, Planckian thermalization $\tau_{\rm eq}\sim \hbar/T$ and causality forbid a Fermi-liquid like resistivity $\rho\propto T^2/E_F$. It is interesting that these arguments tie  Planckian thermalization with linear-in-$T$ resistivity {\em without} the need of a quasiparticle picture and tricky notion of a `transport time' $\tau_{\rm tr}$ extracted from $\rho_{\rm dc}$.

Previous discussions of the Planckian bound were challenged by the need to separate the elastic and inelastic collision rates \cite{Hartnoll:2021ydi,Poniatowski_2021}. Our bound \eqref{eq_bound_conclusion} holds whether or not elastic collisions are present, and whether or not the kinetic theory description necessary to formulate this distinction even applies. However, it still features an incarnation of this distinction that can be formulated without referring to quasiparticles: diffusive hydrodynamics can emerge from disorder or many-body effects. In the former case, there is no limit to how fast diffusion emerges. This is compatible with our bound \eqref{eq_bound_conclusion}, because in this case diffusion will still emerge at the Planckian time in the two-sided correlator (see App.~\ref{app_superplanckianhydro}). It would be interesting to further explore possible superplanckian theories of hydrodynamics that nevertheless satisfy \eqref{eq_bound_conclusion}. The results of Ref.~\cite{Delacretaz:2023pxm} imply that this cannot occur in relativistic quantum field theories, unless the number of local degrees of freedom $N\to\infty$.

Finally, let us comment on the fact that our lower bound on $\tau_{\rm eq}$ in \eqref{eq_bound_simple} scales with the dimension $d$, implying that fastest quantum thermalizers are slower in higher dimensions. Holographic theories allow for one test of this statement. Black holes and black branes have two types of quasi-normal modes: hydrodynamic, with small imaginary frequency at small momenta frequency, and non-hydrodynamic, with finite imaginary frequency at small momenta. Hydrodynamics emerges when the latter have decayed. Our bound suggests that there must be non-hydrodynamic modes with $(\Im \omega) d/T \sim 1$ even for large $d$.
Interestingly, for many black branes it is indeed the case \cite{Morgan:2009pn}.

\subsection*{Acknowledgements}
We are grateful to Sihan Chen, Minjae Cho, Debanjan Chowdhury, Sandeep Joy, Jaewon Kim, Zohar Komargodski, Andrew Lucas, Ruochen Ma, Ruchira Mishra, Nick O'Dea, Zhiyuan Sun, Chong Wang, and especially Akshat Pandey and Amit Vikram for helpful questions and discussions. We are also grateful to the organizers and participants of the 2025 Boulder Summer School, where this work was presented. MQ is supported by the Simons Collaboration on Ultra-Quantum Matter, which is a grant from the Simons Foundation (No. 651442). LD is supported by a NSF CAREER award (DMR-2441227) and a Sloan fellowship.
AM would like to thank C.~King for a moral support.

\appendix

%######################################################################%
%======================================================================%
%======================================================================%
%======================================================================%
%######################################################################%
\section{Planckian bound without $G^S_{nn} > 0$} \label{app:weakening_assumptions}

In the main text, we proved a Planckian bound \eqref{eq_bound_simple} for systems whose symmetric Green's function obeys $G^S_{nn}(t) > 0$. In this appendix, we derive a Planckian bound which does not rely on this condition. Our starting point is the analyticity bound in Eq.~\eqref{eq:bound1}. We define $\tau_{\rm eq}$ to be the earliest time later than $t_0$ such that the relative error between $F_{nn}(t)$ and $F_{\text{hydro}}(t)$ is less than $\epsilon$. Note that this is the same definition as in the main text, with the additional condition that $\tau_{\rm eq} > t_0$.

\begin{figure}
\centering
\begin{overpic}[width=0.85\linewidth,tics=10,trim={-0.0cm 0cm 0.07cm 0cm},clip]{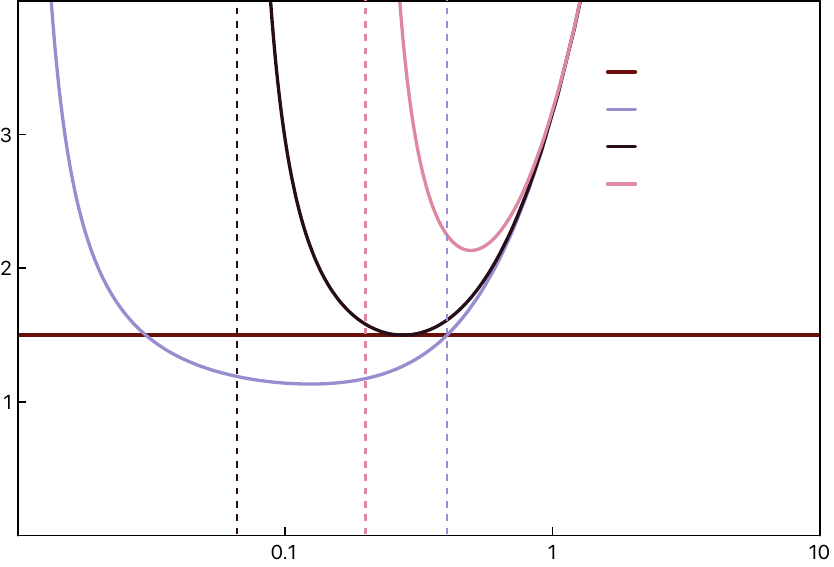}
  %\put (-7,15) {\rotatebox{90}{$\frac{\pi t}{\beta} \coth( \frac{\pi}{\beta}(t-t_0))$}}
	 \put (78,59) {$3/2$} 
	 \put (78,54) {$t_0 = 0.01 \beta$} 
  \put (78,49) {$t_0 = t_c$}
  \put (78,44) {$t_0 = 0.2 \beta$}
  \put (50,-2) {$t / \beta$}
\end{overpic}
\caption{\label{fig:coth_bound}Illustration of the bound $\frac{d}{2} \leq \frac{\pi t}{\beta} \coth (\frac{\pi}{\beta}(t-t_0))$ for different values of $t_0$. Dashed lines indicate bounds on $\tau_{\rm eq}$ at each value of $t_0$. Taking the weakest such bound (black) gives $\tau_{\rm eq} \geq t_c$. }
\end{figure}
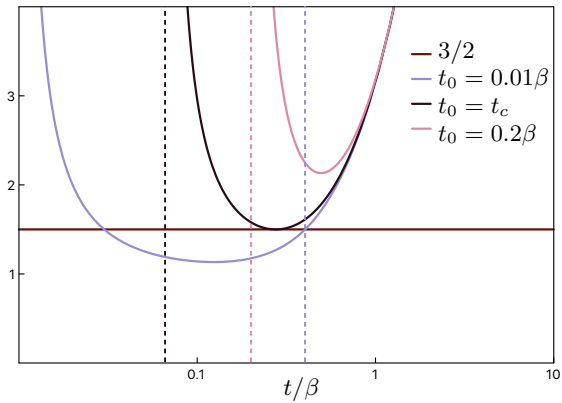

We begin with the case of spatial dimension $d \geq 3$, for which the proof in the main text only needs to be slightly modified. Assume the system has thermalized at a time $t > \tau_{\rm eq}$, so that $F_{nn}$ and its time derivative can be approximated to error $O(\epsilon)$. Because $t > t_0$, the bound \eqref{eq:bound1} gives 
\begin{equation} \label{eq:weakbound}
\frac{d}{2t} \leq \frac{\pi}{\beta} \coth \left( \frac{\pi}{\beta} (t - t_0) \right).
\end{equation}
Let us imagine $t_0$ as a tunable parameter, and consider the family of inequalities obtained by varying $t_0$. 
At large $t$, the right-hand side of \eqref{eq:weakbound} approaches $\pi/\beta$ exponentially quickly, while the left-hand side decays as $1/t$. Therefore, the inequality is always satisfied at sufficiently large $t$ regardless of the value of $t_0$. Above a critical value $t_0 > t_c$, \eqref{eq:weakbound} places no constraint on $t$. Below the critical value $t_0 < t_c$, the inequality imposes a bound $t > t_* > t_c$, where $t_*$ is the largest intersection of the RHS and LHS of \eqref{eq:weakbound}. In each case, the bound imposes the condition $t > t_c$.
The value of $t_c$ is determined by the requirement that the LHS and RHS of \eqref{eq:weakbound} touch each other.
We plot the inequality \eqref{eq:weakbound} for different values of $t_0$ in Fig.~\ref{fig:coth_bound}. We find, for $d=3$, that $t_c = \frac{\sqrt{3}-\log (2+\sqrt{3})}{2\pi} \beta \approx 0.066\, \beta $. We therefore arrive at the Planckian bound 
\begin{equation} \label{eq:3d_planckian_bound}
\tau_{\rm eq} \geq  \frac{\sqrt{3}-\log (2+\sqrt{3})}{2\pi}  \frac{\hbar}{T}
\end{equation}  
for the emergence of diffusion in $d=3$ spatial dimensions. For general $d\geq3 $ the lower bound is
\begin{equation} \label{eq_taueq_d}
\tau_{\rm eq} \geq \frac{\sqrt{d^2-2d} - \log \left(-1+d + \sqrt{d(d-2)} \right)}{2\pi} \frac{\hbar}{T}.
\end{equation}
Like the bound in the main text, this bound on $\tau_{\rm eq}$ scales linearly with $d$.  

This approach is insufficient to prove a Planckian bound for $d=1$ and $d=2$. The reason is that the \eqref{eq:weakbound} is always satisfied for any value of $t_0$, and so fails to provide any constraint on $t$. We can deal with $d=1,2$ by introducing a more comprehensive  definition of $\tau_{\rm eq}$ that incorporates knowledge of the spatial dependence of $F_{nn}(t,x)$: specifically, hydrodynamics should describe this observable at late times for any $|x|\lesssim \sqrt{Dt}$, and not only $x=0$. This will lead to a general Planckian bound on all theories where diffusion emerges, and which can be generalized to other types of hydrodynamic behavior. 

We will apply the analyticity bound \eqref{eq:bound1} to the hydrodynamic correlation function $F_{n(x) n}(t) \equiv F_{nn}(t,x)$. In contrast to the main text, where we set $x=0$, here it will be useful to consider the $x$ dependence of $F_{nn}$. For any $x$, $\Re(F_{nn}(x,z))$ is positive in the hydrodynamic regime $\Re(z) \to \infty$, so there exists a $t_0(x) < \infty$ for each $x$ such that $\Re(F_{nn}(x,z)) > 0$ for all $\Re(z) \geq t_0(x)$. We can therefore apply Eq.~\eqref{eq:bound1} to give a family of inequalities
\begin{equation} \label{eq:x_dependent_inequality}
    \frac{|F'_{nn}(t,x)|}{F_{nn}(t,x)}\leq \frac{\pi}{\beta} \coth \left( \frac{\pi}{\beta} (t-t_0(x)) \right)
\end{equation}
parametrized by $x$. The cutoff time $t_0(x)$ can in general be $x$ dependent. Our definition of $\tau_{\rm eq}$ needs to be modified to take this $x$-dependent cutoff time into account. 

Motivated by this, it is helpful at this stage to recall the scaling limit where diffusive hydrodynamics emerges for thermalizing systems. The hydrodynamic EFT is valid at time scales $t \gtrsim \tau_{\rm eq}$ and at length scales $ x^2 \lesssim Dt$. In particular, the EFT breaks down at scales $x^2 \gg Dt$ \cite{Mishra:2025vlf}, so some care must be taken to avoid extrapolating the EFT beyond its regime of applicability. It is for this reason that we do not simply consider Eq.~\eqref{eq:x_dependent_inequality} for a fixed value of $x$.  

Since hydrodynamics emerges for $Dt \gtrsim x^2$, and $\Re(F_{nn}(x,z))$ is positive in the hydrodynamic regime, hydrodynamics predicts
\begin{equation}
    t_0(x) < \mathcal{O}(x^2).
\end{equation}
If this condition on $t_0(x)$ is violated, then hydrodynamics fails to emerge, and $\tau_{\rm eq} = \infty$. This condition ensures that for any $x$, there always exists a $t = \mathcal{O}(x^2)$ such that $t > t_0(x)$. 

We define the thermalization time $\tau_{\rm eq}$ as follows. Fix an $\mathcal{O}(1)$ dimensionless positive constant $\eta$. Let $x_*$ be the largest solution of $t_0(x) = x^2/(4D\eta)$, and let $t_* \equiv t_0(x_*) = x_*^2/(4D\eta)$. Let $t_0$ denote the maximum value of $t_0(x)$ over all $x < x_*$. We define $\tau_{\rm eq}$ to be the earliest time later than $t_0$ such that the relative error between $F_{nn}(t,x)$ and $F_{cl}(t,x)$ is less than $\epsilon$, for all $t > \tau_{\rm eq}$ and $x^2 \leq 4 D \eta t$. We note that this is a stronger condition than what was required in the main text, since we require that correlation functions agree up to a length scale $x^2 \lesssim Dt$ rather than just at $x = 0$.

With this definition of $\tau_{\rm eq}$ at hand, we can derive a Planckian bound on $\tau_{\rm eq}$ for all spatial dimensions $d$. Fix a $y$ such that  $\eta > d/2+1$. Assume the system has thermalized at a time $t > \tau_{\rm eq}$. Then \eqref{eq:x_dependent_inequality} gives 
\begin{equation}
\begin{aligned}
    \left| - \frac{d}{2t} + \frac{x^2}{4Dt^2} \right| &\leq \frac{\pi}{\beta} \coth \left( \frac{\pi}{\beta} (t - t_0(x) )\right) + \mathcal{O}(\epsilon) \\
    &\leq \frac{\pi}{\beta} \coth \left( \frac{\pi}{\beta} (t - t_0 )\right) + \mathcal{O}(\epsilon)
\end{aligned}   
\end{equation}
for all $x^2 \leq 4 D \eta t$. Specializing to $x^2 = 4D\eta t$, the left-hand side simplifies to
\begin{equation}
    \frac{\eta - d/2}{t} \leq \frac{\pi}{\beta} \coth \left( \frac{\pi}{\beta} (t - t_0 )\right).
\end{equation}
Following a similar procedure to the $d=3$ case, we can combine this inequality with $t > t_0$ to obtain a universal bound 
\begin{equation}
    \tau_{\rm eq} > c_\eta \frac{\hbar}{T}
\end{equation}
where the prefactor $c_\eta$ is given by replacing $d \to 2 \eta - d$ in \eqref{eq_taueq_d}. 
Choosing larger values of $\eta$ means requiring that hydrodynamics emerge over a larger length scale, naturally increasing the time required to thermalize.

%######################################################################%
%======================================================================%
%======================================================================%
%======================================================================%
%######################################################################%
\section{A superplanckian theory of hydrodynamics}\label{app_superplanckianhydro}

The Planckian bound derived in the main text made use of the analyticity properties of the two-sided correlator \eqref{eq:Fnn_def}. 
In this appendix, we explore how this additional condition would constrain a putative ``superplanckian'' theory of hydrodynamics valid at times parametrically smaller that $\beta$.

To illustrate such a theory, consider the following generalization of \eqref{eq_SEFT_ex}:
\begin{align}\label{eq_SEFT_ex}
S[\mu,\phi_a]=&\chi \int \mu (\partial_t + D \nabla^2 g(-\partial_t^2))\phi_a \\ \notag
    &\quad+ iTD \nabla\phi_a \frac{\tfrac12\beta\partial_t}{\tan \tfrac12\beta\partial_t} g(-\partial_t^2)\nabla\phi_a \, .
\end{align}
with an arbitrary function $g\geq 0$. The last term was chosen such that the theory exactly satisfies KMS symmetry \cite{Crossley:2015evo}. This is not the most general possible superplanckian theory of hydrodynamics, but will suffice for our purposes.

We now show that correlation functions of this theory are not a priori subject to the analyticity bounds of Sec.~\ref{sec:analyticity}. The retarded Green's function of the charge density is given by
\begin{equation}
G^R(\omega,q)
	= \chi + \chi\frac{i\omega}{-i\omega + D q^2 g(\omega^2)}
\end{equation}
Combining the standard relation between $G^R$ and $G^S$ with the analyticity of $F$ \eqref{eq_GS_GA_1}, we have
\begin{equation}
    F(\omega, q) = \frac{\chi D q^2 g(\omega^2)}{(D q^2 g(\omega^2))^2 + \omega^2} \frac{\omega}{\sinh \left( \beta \omega/2 \right)} \, .
\end{equation}
Integrating over $q$ to obtain correlation functions at $x=0$ gives
\begin{equation} \label{eq_Fomega}
    F(\omega) =  \frac{\chi \sec \frac{\pi d}{4}}{2^{d+1} \pi^{\frac{d}{2}-1}\Gamma(\frac{d}{2})} \frac{|\omega|^{d/2 - 1}}{(D g(\omega^2))^{d/2}} \frac{\omega}{\sinh \left( \beta \omega/2 \right)} \, .
\end{equation}
Interestingly, if we choose
\begin{equation}
    g(\omega^2) = \left[ \frac{\beta\omega/2}{\sinh(\beta \omega/2)} \right]^{2/d} \, ,
\end{equation}
then $F(\omega, x=0)$ reduces to $F_{\text{hydro}}(\omega, x=0)$ exactly, so that all $1/t$ corrections in \eqref{eq_Fnn_hydro} would vanish, violating our result that hydrodynamics cannot onset arbitrarily early in $F(t)$. Of course, the contradiction comes from the fact that this $F(t)$ obtained from the EFT does not satisfy the analyticity constraints that a two-sided correlator should satisfy in any unitary quantum mechanical system.

How does this additional analyticity condition constrain this ``superplanckian" theory of hydrodynamics? Analyticity of $F(t)$ in the strip implies $F(\omega)$ must decay at least as $e^{- \beta \omega/2}$ at large frequency. This large frequency constraint goes beyond constraints usually imposed on hydrodynamic EFTs, which are expansions in low frequency and wavevector. Analyticity of $F(t)$ thus implies that $g(\omega^2)$ cannot decay too fast at large frequency:
\begin{equation}
    g(\omega^2) \geq e^{- c \omega} \quad \text{as $\omega\to \infty$}\, , \quad \text{ for any $c > 0$}. 
\end{equation}

To obtain an example of a correlation function satisfying all analyticity conditions, and exhibiting diffusion at late times, let us consider $g=1$. One then finds
 \begin{equation}
 \begin{split}
 F(t)
 	&= \frac{\chi d/2}{(4\pi D)^{d/2}} \Gamma(-\tfrac{d}2) \int_{-\infty}^{\infty} \frac{d\omega}{2\pi} e^{-i\omega t}  \frac{(-i\omega)^{d/2} - (i\omega)^{d/2}}{2i \sinh \frac{\beta \omega}{2}}\\
 	&= \frac{\chi d/4}{(4\pi D)^{d/2}} \frac{\sec \frac{d\pi}4 }{\beta^{\frac{d}2+1}} \left[\zeta \left(\frac{d}2 + 1, \frac12 + \frac{i t}\beta\right) + (t\to -t)\right]\\
 \end{split}
 \end{equation}
The integral was obtained by contour integration, and involves the Hurwitz zeta function. The result is manifestly KMS invariant $F(t) = F(-t)$, and is analytic in the expected strip (it has branch points for $t = i\beta (n+\frac12)$ with $n\in \mathbb Z$). It is diffusive at late times: using $\zeta(1+\frac{d}2,x)\to \frac{2/d}{x^{d/2}}$ as $x\to \infty$ one finds as expected
\begin{equation}
F(t)
	\simeq \frac{\chi}{(4\pi Dt)^{d/2}} \frac{1}{\beta}\, .
\end{equation}
This result for the two-sided correlator, together with $F_{\text{hydro}}$, is plotted in Fig.~\ref{fig_Fig1}. This correlator obeys the bound \eqref{eq_FprimeOverF}, even though $G^S(t) > 0$ is not satisfied by this ``superplanckian" model of hydrodynamics.

%######################################################################%
%======================================================================%
%======================================================================%
%======================================================================%
%######################################################################%
\section{A bound on the correlator}
\label{app:lattice_bound}
The correlator at hand is
\begin{equation}
   \frac{1}{\Tr(y_+ y_-)} \Tr(y_+ A y_- B), \ y_{\pm} = e^{H(-\beta/2 \pm \tau)}. 
\end{equation}
First, by the virtue of the Cauchy--Schwartz inequality we can instead consider a diagonal correlator:
\begin{equation}
    f = \frac{1}{Z} \Tr(y_- A y_+ A).
\end{equation}
We can write it in the energy basis as
\beq
\label{eq:y_sum}
\frac{1}{Z}\sum_i (y_-)_i (y_+)_j |A_{ij}|^2.
\eeq
Matrix $M_{ij} = |A_{ij}|^2$ is symmetric and doubly sub-stochastic (sum of the entries in each row/column is less than 1):
\beq
\sum_i M_{ij} \le 1, \sum_j M_{ij} \le 1,
\eeq
because
\beq
\sum_i M_{ij} = \sum_i \langle i| A| j \rangle \langle j | A| i \rangle = \langle j | A^2 | j \rangle \le ||A||^2_\infty.
\eeq
We can add extra positive numbers to the diagonals of $M_{ij}$, which will only increase the sum (\ref{eq:y_sum}). We can do it to make it doubly-stochastic:
\beq
\sum_i M_{ij} = 1, \sum_j M_{ij} = 1.
\eeq
Then according to Darboux theorem,
new $M$ can be represented as a convex combination of permutation matrices:
\beq
M = \sum_\alpha q_\alpha P_\alpha, \ q_\alpha >0, \ \sum_\alpha q_\alpha=1.
\eeq
Let us look at $y_- P_\alpha y_+$.
Since $y_\pm$ have the same ordering, $(y_+)_1 > (y_+)_2 > \dots$ (here $|\tau|<\beta/2$ is important), the rearrangement lemma\footnote{The lemma says that if we have two ordered sequences of positive numbers $p_i, q_i$ and we take their dot product after permuting one of them, 
$$
\sum_i p_i q_{\sigma(i)}
$$
then this sum is maximized by the trivial permutation $\sigma$. 
} says that 
\beq
\frac{1}{Z} y_- P_\alpha y_+ \le \frac{1}{Z} \sum_i (y_-)_i (y_+)_i = \frac{1}{Z}\Tr e^{-\beta H}=1.
\eeq
This concludes the proof.

\bibliography{biblio}

\end{document}